\documentstyle[12pt]{article}
\input{epsfig.sty}
\newcommand{\beq}{\begin{eqnarray}}
\newcommand{\eeq}{\end{eqnarray}}
\begin{document}
\title{$\Psi(2S)$ and $\Upsilon(3S)$ Suppression in p-Pb 8 TeV Collisions and 
Mixed Heavy Quark Hybrid Mesons}
\author{Leonard S. Kisslinger$^{1}$\\
Department of Physics, Carnegie Mellon University, Pittsburgh PA 15213 USA.\\
Debasish Das$^{2,3}$\\
High Energy Nuclear and Particle Physics Division,\\
 Saha Institute of Nuclear 
Physics, 1/AF, Bidhan Nagar, Kolkata 700064, INDIA.}
\date{}
\maketitle
\vspace{-1cm}

\noindent
1) kissling$@$andrew.cmu.edu \hspace{1cm} 2)dev.deba$@$gmail.com; 
3) debasish.das@saha.ac.in

\begin{abstract} This brief report is an extension of a previus publication
on $\Psi(2S)$ to $J/\Psi(1S)$ suppression in p-Pb collisions at 5.02 TeV
to estimate $\Psi(2S)$ to $J/\Psi(1S)$ and $\Upsilon(3S)$ to $\Upsilon(1S)$ 
suppression via p-Pb collisions at 8 TeV as proposed by the LHCb.
\end{abstract}

\noindent
PACS Indices:12.38.Aw,13.60.Le,14.40.Lb,14.40.Nd

\section{Introduction}

  Recently $\Psi(2S)$ suppression $S_A=[\sigma_{\Psi(2S)}/\sigma_{J/\psi}]_{pPb}/
[\sigma_{\Psi(2S)}/\sigma_{J/\psi}]_{pp}$ in p-Pb collisions
at E= 5.02 TeV, and $\Upsilon(3S)$ suppression 
$S_A=[\sigma_{\Upsilon(3S)}/\sigma_{\Upsilon(1S)}]_{PbPb}/
[\sigma_{\Upsilon(3S)}/\sigma_{\Upsilon(1S)}]_{pp}$ in Pb-Pb collisions, were 
estimated\cite{lsk16} and compared to experimental\cite{alice14a,alice14b}. 
By using the mixed 
heavy quark hybrid theory\cite{lsk09} it was found that the theoretical 
estimates agreed with experiments within errors in experiments and 
theory. This was reviewed\cite{lskdd16} before Ref.\cite{lsk16} was published.

Since the article Ref.\cite{lsk16} was published  LHCb has recently
submitted a document\cite{lhcb16} to support the request for p-Pb and Pb-p 
collisions at 8 TeV to produce $\Psi$ and $\Upsilon$ mesons to measure 
``cold nuclear matter effects'', which we call $\Psi(2S)/(J/\Psi)$ and 
$\Upsilon(3S)/\Upsilon(1S)$ suppression.
The objective of the present work is to estimate  $J/\Psi(1S)$, $\Psi(2S)$,
$\Upsilon(1S)$, and $\Upsilon(3S)$  suppression using the mixed heavy quark 
theory for p-Pb vs p-p collisions in anticipation of the future LHCb 
experiment\cite{lhcb16}.

Three decades ago $J/\Psi$ suppression by the formation of the Quark-Gluon
Plasma (QGP) in Reltivistic Heavy Ion Collisions (RHIC) was 
estimated\cite{ms86}. This is closely related to using the mixed hybrid theory 
to detect the formation of the QGP via RHIC\cite{klm14}, since gluons in the
QGP enhance the production of  $\Psi(2S)$ and $\Upsilon(3S)$ states due to
the active gluon in their hybrid component. This is reviewed in the 
following section.

  In the present work the on $\Psi(2S)$, $J/\Psi(1S)$, $\Upsilon(1S)$, and 
$\Upsilon(3S)$  suppression produced in p-Pb collisions at 8 TeV we employ 
the theoretical methods of Ref.\cite{lsk16} using scenario 1. of 
Ref\cite{spi99}, to estimate $S_A=[\sigma_{\Psi(2S)}/\sigma_{J/\psi}]_{pPb}/ 
[\sigma_{\Psi(2S)}/\sigma_{J/\psi}]_{pp}$, $S_A=[\sigma_{\Upsilon(3S)}/
\sigma_{\Upsilon(1S)}]_{pPb}/[\sigma_{\Upsilon(3S)}/\sigma_{\Upsilon(1S)}]_{pp}$, in p-Pb 
8 TeV collisions.

\newpage
\section{Theoretical $\Psi(2S)$ to  $J/\Psi(1S)$ and $\Upsilon(3S)$
to $\Upsilon(1S)$ suppression in p-Pb collisions and the mixed 
heavy quark hybrid theory}

In this section we start our theortical estimate of the suppression, $S_A$, of 
charmonium and bottomonium states by a brief review of the mixed hybrid theory,
followed by the theoretical estimate of $S_A$ for $\Psi(2S)$, $J/\Psi(1S)$,
$\Upsilon(3S)$, and $\Upsilon(1S)$ in p-Pb collisions.

\subsection{Review of Mixed Heavy Hybrid States via QCD Sum Rules }

The nature of the $J/\Psi(1S)$, $\Psi(2S)$ as standard, hybrid, or mixed 
standard-hybrid charmonium states, and $\Upsilon(1S)$, $\Upsilon(3S)$ as 
standard, hybrid, or mixed standard-hybrid bottominium states was studied in, 
Ref.\cite{lsk09} making use of QCD Sum Rules\cite{sz79}.

 The operator that produces the mixed charmonium and hybrid 
charmonium states, with $f$ determined from the Sum Rule, is
\beq
\label{11}
        J_{C-HC} &=& f J_H + \sqrt{1-f^2} J_{HH} \; ,
\eeq
with $J_H|0> = |c\bar{c}(0)>, J_{HH}|0> = |[c\bar{c}(8)g](0)>$,
where $|c\bar{c}(0)>$ is a standard Charmonium state, while a hybrid 
Charmonium state $|[c\bar{c}(8)g](0)>$ has $c\bar{c}(8)$ with color=8 and a 
gluon with color=8. In Ref.\cite{lsk09} it was found that for $J/\Psi(1S)$
$f^2 \simeq 1.0$ while for $\Psi(2S)$ $f^2 \simeq 0.5$. Thus
\beq
\label{mixedcharmhybrid}
    |J/\Psi> &\simeq& |c\bar{c}(0)(1S)> \nonumber \\
    |\Psi(2S)> &\simeq& 0.7071|c\bar{c}(0)(2S)> +0.7071|[c\bar{c}(8)g](0)(2S)> 
\; ,
\eeq
therefore the $\Psi(2S)$ meson is 50\% normal charmonium and 50\% hybrid 
charmonium, while the $J/\Psi$ is a normal charmonium meson.

Using a similar QCD Sum Rule calculation for bottominium states\cite{lsk09}
it was found that the $\Upsilon(1S)$ is a standard bottominium meson, while
the $\Upsilon(3S)$ is 50\% normal  and 50\% hybrid bottominium meson:
\beq
\label{mixedbottomhybrid}
    |\Upsilon(1S)> &\simeq& |b\bar{b}(0)(1S)> \nonumber \\
    |\Upsilon(3S)> &\simeq& 0.7071|b\bar{b}(0)(3S)> +0.7071|[b\bar{b}(8)g](0)
(3S)> \; .
\eeq

We shall use this to estimate the ratios of suppression of  $\Psi(2S)$ 
to $J/\Psi$ and  $\Upsilon(3S)$ to  $\Upsilon(1S)$ is in p-Pb collisions.
 
\subsection{Theoretical $\Psi(2S)$ to  $J/\Psi(1S)$ suppression in p-Pb
collisions}

The suppression, $S_A$, of a charmonium state is given by the interaction with
nucleons as it traverses the nucleus. In this subsection we give a brief
review of $S_A$ for standard and hybrid charmonium mesons derived in 
Ref.\cite{lsk16}.

\beq
\label{SA}
    S_A &=& e^{-n_o\sigma_{\Phi N} L} \; ,
\eeq
where $\Phi$ is a $c\bar{c}$ or $c\bar{c}g$ meson,  $L$ is the length of 
the path of $\Phi$ in nuclear matter $\simeq$ 8 to 10 fm for
p-Pb collisions, with nuclear matter density 
$n_o=.017 fm^{-3}$, and $\sigma_{\Phi N}$ is the cross section for $\Phi$-
nucleon collisions.

  From Refs.\cite{ks96} the cross section for standard charmonium $c\bar{c}$
 meson via strong QCD interactions with nucleons is given by
\beq
\label{sigma-c-barc} 
   \sigma_{c \bar{c} N} &=& 2.4 \alpha_s \pi r_{c \bar{c}}^2 \; ,
\eeq
where  $\alpha_s\simeq $ 0.118, and the
charmonium meson radius $r_{c \bar{c}} \simeq \not\!h/(2 M_c c)$, with $M_c$ 
the charm quark mass. Using $2 M_c \simeq M_{J/\Psi} \simeq $ 3 GeV,
$4 r_{c \bar{c}} \simeq 6 \times 10^{-17}m= 0.06 fm$. Therefore,
$\sigma_{c \bar{c} N} \simeq 3.2 \times 10^{-3} fm^2=3.2 \times 10^{-2}$ mb.

  Taking $L \simeq$ 8-10 fm and $n_o=.017 fm^{-3}$
\beq
\label{SAcc}
        n_o\sigma_{c \bar{c} N} L &\simeq& 0.0022 \nonumber \\
        S_A^{c\bar{c}} &=& e^{- n_o\sigma_{c \bar{c} N} L} \simeq 1.0 \; .
\eeq
 
 On the other hand, the cross section for hybrid charmonium $c\bar{c}g$ meson 
via strong QCD interactions with nucleons has been estimated in Ref\cite{ks96}
as $\sigma_{c\bar{c}g N} \simeq$ 6-7 mb. In the present work we use
\beq
\label{sigmacbarcg}
     \sigma_{c\bar{c}g N} &\simeq & 6.5 mb \; ,
\eeq
with the result
\beq
\label{SAccg}
      n_o\sigma_{c\bar{c}g N} L&\simeq& 0.88 {\rm \;to\;}1.1 \nonumber \\
   S_A^{c\bar{c}g} &\simeq& 0.4 {\rm \;to\;} 0.33  \; .
\eeq

Using the mixed hybrid model one finds for the ratio of $\Psi(2S)$ to  
$J/\Psi(1S)$ suppression in p-Pb collisions
\beq
\label{SAratio}
    R^{\Psi(2S)/(J/\Psi)}|_{theory}&\simeq& \frac{1+0.4{\rm \;to\;} 0.33}{2} 
\nonumber \\
          &\simeq& 0.66{\rm \;to\;}0.7 \; .
\eeq  

As pointed out in Ref.\cite{alice15}, the Color Glass Condensate 
model\cite{fw13} overestimates the suppression, while other theoretical  models
successfully estimate  $J/\Psi$ suppression, but do not treat the $\Psi(2S)$
suppression. Since they would use a standard 
$c\bar{c}$ rather than the mixed hybrid theory with a $c\bar{c}g$ component
as in the present work, they would probably underestimate the $\Psi(2S)$
suppression.

\subsection{Theoretical $\Upsilon(3S)$ to  $\Upsilon(1S)$ 
suppression in p-Pb collisions}

For a standard bottomonium meson state  or a hybrid bottomonium
state the equation for suppression is given by 
Eq({\ref{SA}) where $\Phi$ is a $b\bar{b}$ or $b\bar{b}g$ meson. As for
charmonium  $L$ is the length of  $\simeq$ 8 to 10 fm for
p-Pb collisions, with nuclear matter density $n_o=.017 fm^{-3}$. 

 From Eq(\ref{sigma-c-barc}) the cross section for standard bottomonium
 $b\bar{b}$ meson via strong QCD interactions with nucleons,
$\sigma_{b \bar{b} N} = 2.4 \alpha_s \pi r_{b \bar{b}}^2$ differs from
$\sigma_{c \bar{c} N}$ by a factor of $ M_c^2/M_b^2\simeq 0.09$. Therefore,
$\sigma_{b \bar{b} N} \simeq 0.09 \times 3.2 \times 10^{-2} mb \simeq 
3 \times 10^{-3} mb$
\beq
\label{SAcc}
        n_o\sigma_{b \bar{b} N} L &\simeq& 0.0002 \nonumber \\
        S_A^{c\bar{c}} &=& e^{- n_o\sigma_{c \bar{c} N} L} \simeq 1.0 \; .
\eeq

 Similarly, the cross section for hybrid bottomonium $b\bar{b}g$ meson 
via strong QCD interactions with nucleons $\sigma_{b\bar{b}g N}=
\sigma_{c\bar{c}g N}(M_c/M_b)^2\simeq 0.09 \sigma_{c\bar{c}g N} \simeq $ 0.59 mb.
Therefore,
\beq
\label{SAbbg}
      n_o\sigma_{b\bar{b}g N} L&\simeq& 0.08 {\rm \;to\;}0.1 \nonumber \\
   S_A^{b\bar{b}g} &\simeq& 0.92 {\rm \;to\;} 0.90  \; .
\eeq

Using the mixed hybrid model one finds for the ratio of  $\Upsilon(3S)$ to  
$\Upsilon(1S)$  suppression in p-Pb collisions
\beq
\label{SAratio}
    R^{\Upsilon(3S)/\Upsilon(1S))}|_{theory}&\simeq& \frac{1+0.90{\rm \;to\;} 
0.92}{2} 
\nonumber \\
          &\simeq& 0.95{\rm \;to\;}0.96 \; .
\eeq  

\section{Conclusions}

Using the mixed heavy quark hybrid theory for $\Psi(2S)$ and $\Upsilon(3S)$
states we have estimated $\Psi(2S)$ to $J/\Psi$ and $\Upsilon(3S)$ to
$\Upsilon(1S)$ suppression ratios, $R^{\Psi(2S)/(J/\Psi)}|_{theory}$ and
$R^{\Upsilon(3S)/\Upsilon(1S))}|_{theory}$, for p-Pb vs p-p collisions at 8 TeV 
in anticipation of future LHCb experiments\cite{lhcb16}. Note that previous
experiments\cite{alice14a} measuring $R^{\Psi(2S)/(J/\Psi)}$ used  p-Pb collisions 
at 5.02 TeV.

If future experiments measure the $R^{\Psi(2S)/(J/\Psi)}$ and  
$R^{\Upsilon(3S)/\Upsilon(1S))}$ ratios for p-Pb vs p-p collisions, with the 
$\Psi(2S)$ meson predicted to be 50\% normal charmonium and 50\% hybrid 
charmonium and the $\Upsilon(3S)$  meson predicted to 
be 50\% normal bottomonium and 50\% hybrid bottomonium\cite{lsk09}, one would
have another test of the mixed heavy quark hybrid theory.
\vspace{5mm}

\Large
{\bf Acknowledgements}

\normalsize
\vspace{5mm}

Author D.D. acknowledges the facilities of Saha Institute of Nuclear Physics, 
Kolkata, India. Author L.S.K. acknowledges support from the P25 group at Los 
Alamos National Laboratory.

\end{document}